\documentclass{aa}
\def\e{et~al.\ }

\def\r{\mbox{R\,CrB}}
\newcommand{\be}{\begin{equation}}
\newcommand{\ee}{\end{equation}}
\newcommand{\bdm}{\begin{displaymath}}
\newcommand{\edm}{\end{displaymath}}
\usepackage{graphicx}
\begin{document}

   \title{$UBVJHKLM$ photometry and modeling of R Coronae Borealis}

\author{B.F.\,Yudin\inst1\fnmsep\inst2\fnmsep\inst3, J.D.\,Fernie\inst4,
N.R.\,Ikhsanov\inst2\fnmsep\inst5, V.I.\,Shenavrin\inst1,
G.\,Weigelt\inst2}

  \offprints{B. Yudin \\ \email{yudin@sai.msu.ru}}

   \institute{Sternberg Astronomical Institute, Universitetskii
              prospect 13, 119899 Moscow, Russia
              \and
              Max-Planck-Institut f\"ur Radioastronomie, Auf dem
              H\"ugel 69, D-53121 Bonn, Germany
              \and
              Isaac Newton Institute of Chile, Moscow Branch, Russia
              \and
              David Dunlap Observatory, University of Toronto,
              Box 360, Richmond Hill, Ontario L4C 4Y6, Canada
              \and
              Central Astronomical Observatory RAS at Pulkovo,
              Pulkovo 65/1, 196140 St.\,Petersburg, Russia}

   \date{Received ; accepted }

\authorrunning{B. Yudin \e}

\abstract{We present the results of $UBVJHKLM$ photometry of \r\
spanning the period from 1976 to 2001. Studies of the optical
light curve have shown no evidence of any stable harmonics in the
variations of the stellar emission. In the $L$ band we found
semi-regular oscillations with the two main periods of $\sim
3.3$\,yr and $\sim 11.9$\,yr and the full amplitude of $\sim
0\fm8$ and $\sim 0\fm6$, respectively. The colors of the warm dust
shell (resolved by Ohnaka \e \cite{ohnaka01}) are found to be
remarkably stable in contrast to its brightness. This indicates
that the inner radius is a constant, time-independent
characteristic of the dust shell. The observed behavior of the IR
light curve is mainly caused by the variation of the optical
thickness of the dust shell within the interval $\tau(V)=
0.2-0.4$. Anticorrelated changes of the optical brightness (in
particular with $P \approx 3.3$\,yr) have not been found. Their
absence suggests that the stellar wind of \r\ deviates from
spherical symmetry. The light curves suggest that the stellar wind
is variable. The variability of the stellar wind and the creation
of dust clouds may be caused by some kind of activity on the
stellar surface. With some time lag, periods of increased
mass-loss cause an increase in the dust formation rate at the
inner boundary of the extended dust shell and an increase in its
IR brightness. We have derived the following parameters of the
dust shell (at mean brightness) by radiative transfer modeling:
inner dust shell radius $r_{\rm in} \approx 110\,R_*$, temperature
$T_{\rm dust}(r_{\rm in}) \approx 860$\,K, dust density $\rho_{\rm
dust}(r_{\rm in}) \approx 1.1\times 10^{-20}\,{\rm g\,cm^{-3}}$,
optical depth  $\tau(V) \approx 0.32$ at 0.55\,$\mu$m, mean dust
formation rate $\dot{M}_{\rm dust} \approx 3.1 \times
10^{-9}\,{\rm M_{\sun}\,yr^{-1}}$, mass-loss rate $\dot{M}_{\rm
gas} \approx 2.1 \times 10^{-7}\,{\rm M_{\sun}\,yr^{-1}}$, size of
the amorphous carbon grains $\la 0.01\,\mu$m, and $B-V \approx
-0.28$. }

   \maketitle

   \section{Introduction}

R Coronae Borealis is the prototype of a small group of yellow
supergiants (about 35 known members in our Galaxy) characterized
by sudden declines in their optical brightness and extremely
hydrogen-deficient, carbon-rich atmospheres. The visual light
curve of \r\ contains quasi-regular low-amplitude variations
without any dominating periods. These are commonly interpreted in
terms of stellar pulsations and episodic deep declines, which
typically last a few months and have amplitudes of up to
8\,magnitudes. Such events are thought to be the result of
obscuration by dust clouds generated spasmodically and blown away
from the star by radiation pressure (Clayton \cite{clay96} and
references therein).

The near-infrared excess, discovered by Stein \e (\cite{stein69}),
is assigned to a warm dust shell (blackbody temperature of $\sim
900$\,K). This extended dust shell was resolved for the first time
by speckle interferometric observations with the SAO 6 m telescope
(Ohnaka \e {\cite{ohnaka01}). The IR excess contributes about 30\%
of the total flux and is permanently present, regardless of the
visual brightness of the object. The $L$ flux of \r, which is
mainly due to dust emission, varies semi-regularly with a period
of 1260\,days (Feast \e \cite{feast97} and references therein).
The analysis of IRAS observations at 60\,$\mu$m and 100\,$\mu$m
(Gillett \e \cite{gill86}) led to the discovery of a very extended
``fossil'' shell around \r, whose diameter and temperature are
$\simeq 18\arcmin$ and 30\,K, respectively.

In this paper we focus on the investigation of the time-dependent
characteristics of the extended warm dust shell and their
interpretation. In the next section we present the results of our
photometric observations of \r\ ($UBV$ in 1994--1999 and $JHKLM$
in 1983--2001) and the complete tables of photometric observations
of the star in 1976--2001. The basic characteristics of the
optical and IR radiation of the star during its bright state are
analyzed in Sect.\,3. Radiative transfer modeling of the warm dust
shell at its average brightness is described in Sect.\,4. The
variations of the dust shell brightness and colors, and,
correspondingly, its structural parameters, are investigated in
Sect.\,5. Our results are discussed in Sect.\,6.

   \section{Observations}

Our $UBV$ observations of \r\ were carried out in the period
1994--1999 with the 0.25 m telescope of the Fairborn observatory
(Genet \e \cite{genet87}). The results of these observations
together with those of earlier observations (1985--1993) by Fernie
\& Seager (\cite{fernie94}) are summarized in Table\,1 in
\mbox{http://infra.sai.msu.ru/ftp/rcrb}. This table contains 1170
measurements with the accuracy of $0\fm02$ or better. The standard
star was HD\,141352: $V=7.476$, $B-V=0.439$ and $U-B=-0.003$.

$JHKLM$ observations were carried out between 1983 and 2001 with
the 1.25 m telescope of the Crimean Station of the Sternberg
Astronomical Institute (for technical details see Nadjip \e
\cite{nadg86}). The results our observations, together with those
obtained in 1976--1979 (Shenavrin \e \cite{shen79}), are
summarized in Table\,2 in  ${\rm
http://infra.sai.msu.ru/ftp/rcrb}$. This table contains 252
measurements with the accuracy of $0\fm03$ or better. The standard
star was BS\,5947: $J=2.09$, $H=1.60$, $K=1.30$, $L=1.12$ and
$M=1.35$.
    \begin{figure*}

      \resizebox{18cm}{20cm}{\includegraphics{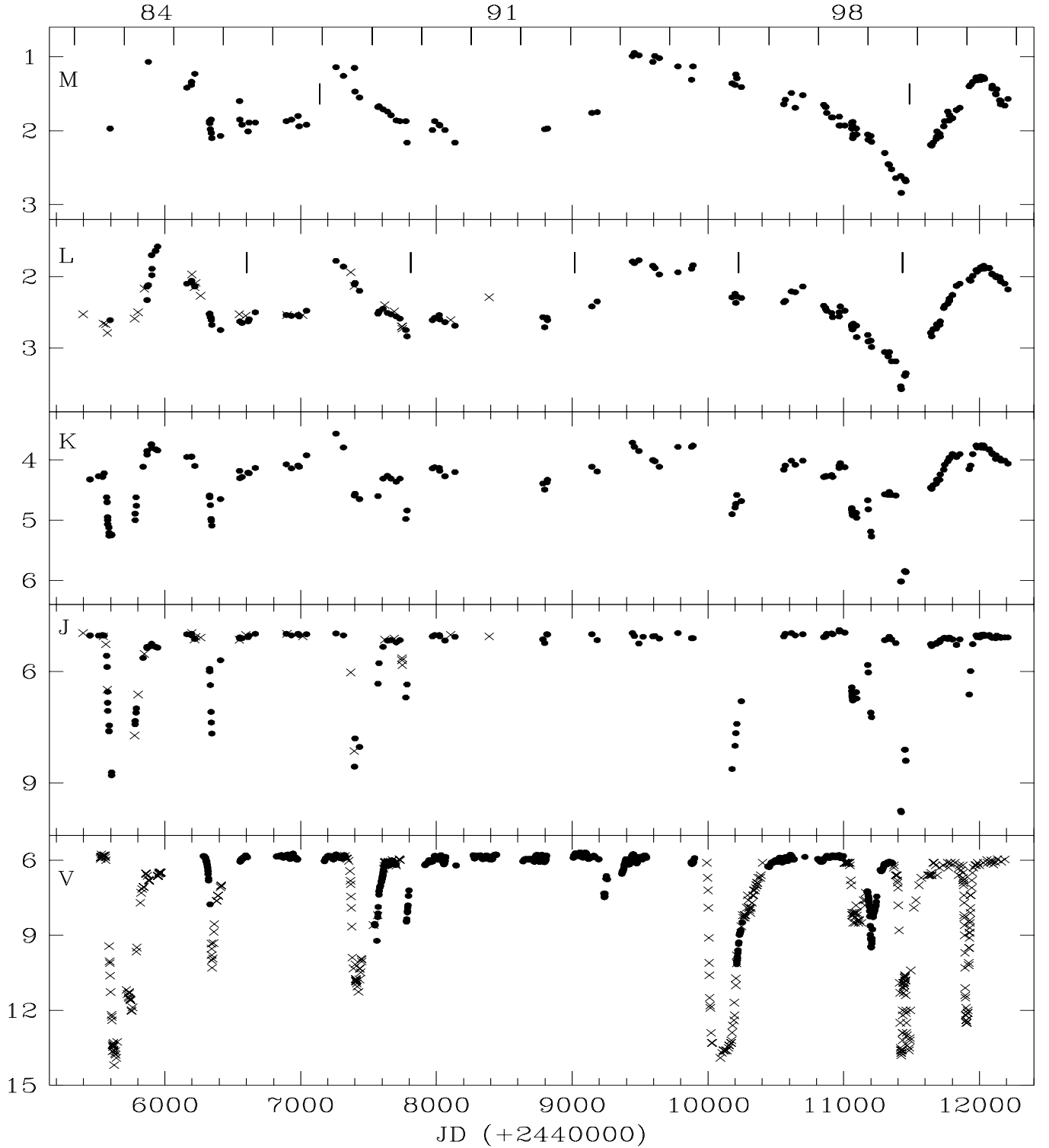}}
        \caption{$VJKLM$ light curves of \r. The dots represent the
observations reported in this paper. The crosses in the visual
light curve denote the data of VSOLJ
(\mbox{http://www.kusastro.kyoto-u.ac.jp/vsnet}), and in the $J$
and $L$ light curves, the data reported by Feast \e
(\cite{feast97}). The vertical bars in the $L$ and $M$ panels mark
the times of the minima calculated from Eqs.~(\ref{m14}) and
(\ref{m15}), respectively.}
      \label{f1}
    \end{figure*}

The $VJKLM$ light curves of \r\ during the period 1983--2001 are
presented in Fig.\,\ref{f1}. Also plotted are a few estimates in
the $V$ band taken from the internet page of the association of
amateur astronomers VSOLJ
(\mbox{http://www.kusastro.kyoto-u.ac.jp/vsnet}) and the
measurements reported by Feast \e (\cite{feast97}) in the $J$ and
$L$ bands. As one can see in this figure, a number of deep
declines of the star's visual brightness were observed in
1983--1989 and 1995--2001, while during the interval 1989--1995
the $V$ flux was almost constant. The very different behavior of
the $L$  and $M$ light curves will be discussed in subsequent
sections.

   \section{ Basic characteristics of R CrB in the bright state}

We call the state of \r\ as bright, if $V \leq 6.05$ and $J \leq
5.25$. In this state the contribution of the dust to the optical
radiation of \r\ is negligible. The average $UBVJHKLM$ and $RI$
magnitudes of \r\ in the bright state are summarized in
Table\,\ref{t3}. These values were derived from Tables\,1\,and\,2,
and the data presented by Fernie (\cite{fernie82}) and Fernie \e
(\cite{fernie86}). The maximum magnitudes of \r\ during the whole
observing period are $V \approx 5.69$ and $J \approx 4.89$.

\begin{table}[t]
 \caption[]{The average magnitudes of \r\ in bright state.
$<m>$, $<m>_{\rm L, max}$, and $<m>_{\rm L, min}$ denote the
magnitudes during the average, maximum and minimum star brightness
in the $L$ band}
   \label{t3}
\begin{tabular}{lcccr}
  \noalign{\smallskip}
  \hline
  \noalign{\smallskip}

Bands & $<m>$ & $<m>_{\rm L, max}$ & $<m>_{\rm L, min}$\\
  \noalign{\smallskip}
  \hline
  \noalign{\smallskip}
$U$ & 6.55 & 6.38 & 6.82  \\
$B$ & 6.49 & 6.35 & 6.70 \\
$V$ & 5.91 & 5.81 & 6.08 \\
$R$ & 5.49 & 5.40 & 5.62 \\
$I$ & 5.31 & 5.23 & 5.40 \\
$J$ & 5.08 & 5.01 & 5.11 \\
$H$ & 4.91 & 4.76 & 5.08 \\
$K$ & 4.13 & 3.74 & 4.56 \\
$L$ & 2.38 & 1.79 & 3.12 \\
$M$ & 1.70 & 0.97 & 2.48 \\
  \noalign{\smallskip}
  \hline
\end{tabular}
\newline
\end{table}

   \subsection{Periodicity}

The Fourier analysis of the visual light curve (1985--1999) during
the bright state has revealed a number of harmonics with the full
amplitude of $\sim 0\fm05$ concentrated in the interval of
30--60\,days. The maximum amplitude of $0\fm07$ has the mode with
the period $P \approx 39.8$\,days (see Fig.\,\ref{f2}). The
corresponding harmonic with the full amplitude of $\sim 0\fm025$
was also found from the analysis of the $B-V$ curve. The
brightness maxima follow the color minima with a lag of 3\,days,
and the ratio of the amplitudes of $(B-V)$ and $V$ harmonics is
about 0.3. No trend has been found in the optical brightness. It
should be noted that on a time-scale of several hundreds of days
\r\ can occasionally exhibit well-developed pulsations with
periods of 35, 44 and 51 days (Fernie \& Seager \cite{fernie94}
and references therein).

   \begin{figure}
     \resizebox{8.5cm}{14cm}{\includegraphics{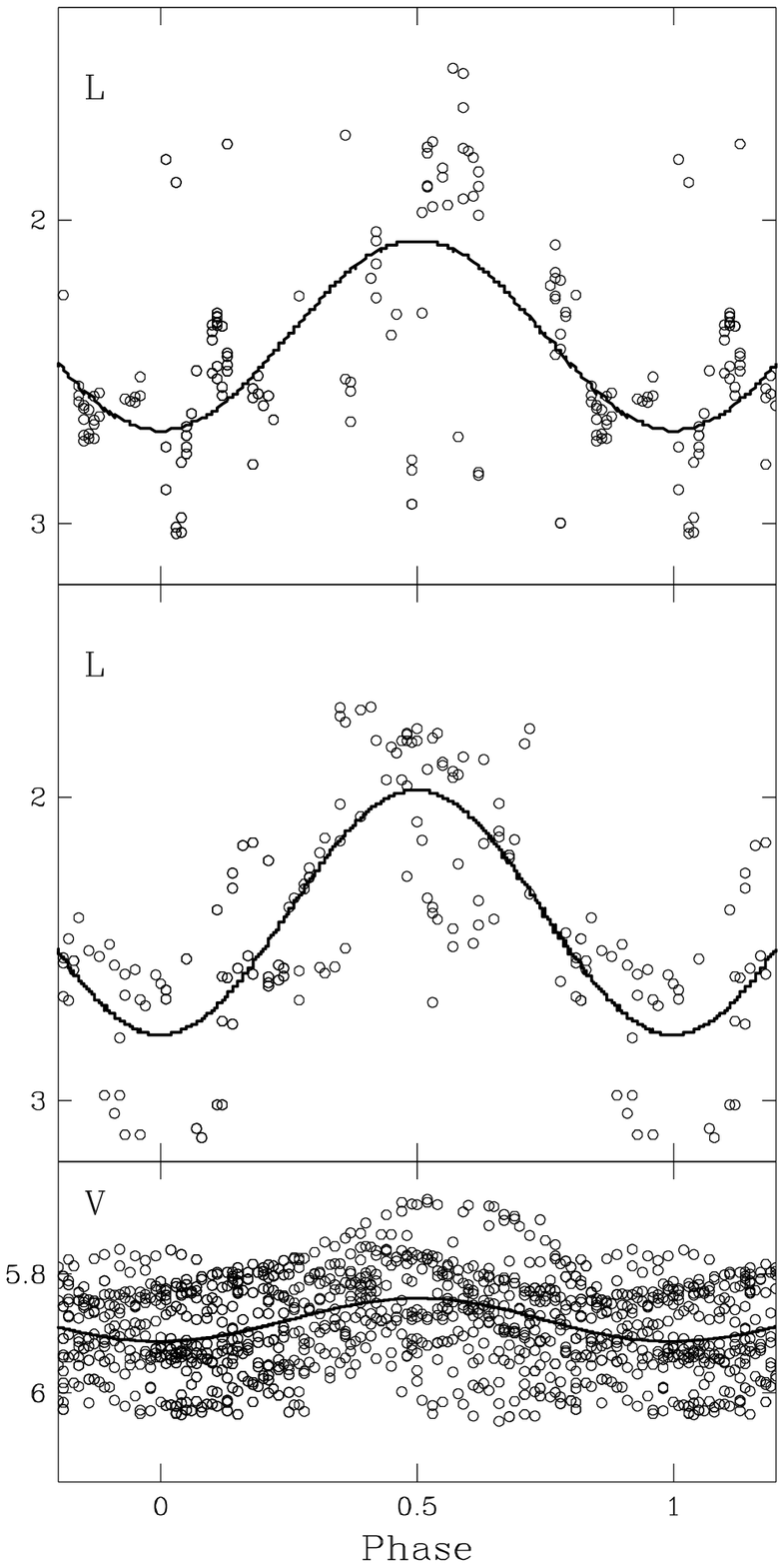}}
        \caption{The $V$ and $L$ light curves of \r\
folded with the periods 39.82, 1206 and 4342\,days, from bottom to
top}.
    \label{f2}
  \end{figure}

The low-amplitude quasi-regular variability in the $J$ band
resembles that in the optical region. A period search has revealed
three main harmonics with periods of 43.7, 33.2, and 28.1\,days
and an amplitude of about $0\fm08$. At the same time Rao \e
(\cite{rao99}) have found that the radial velocity curve exhibits
the only one stable pulsation period of about 42.7\,days.

   \subsection{Spectral energy distribution}

The spectral energy distribution (SED) of \r\ relating to its
average brightness in bright state (see Table\,\ref{t3}) is shown in
Fig.\,\ref{f3}.

   \begin{figure}
     \resizebox{8.5cm}{10cm}{\includegraphics{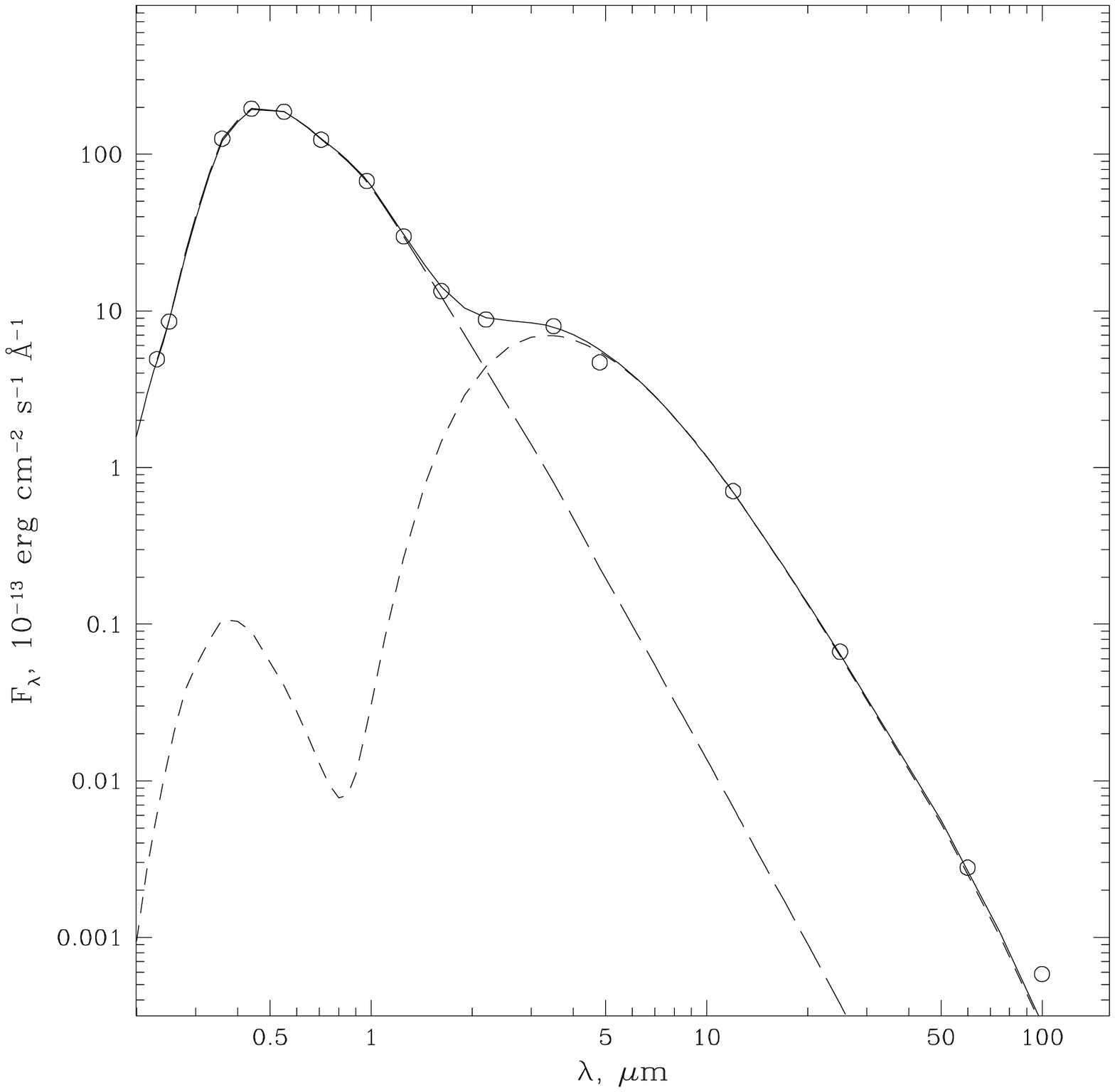}}
        \caption{SED of \r\ for average brightness in bright state
(circles). Long- and short-dashed lines denote the SED of the
star and the dust shell, respectively. The solid line is the
fitted SED (see text)}.
       \label{f3}
    \end{figure}

The stellar magnitudes were corrected for the interstellar
reddening according to the color excess $E(B-V) = 0.05$ (Asplund
\e \cite{asp97}). The flux at 0.23 and 0.25\,$\mu$m was measured
with IUE on 31.05.1991 (see INES data from the IUE satellite),
when the star brightness in the visual and near IR was close to
its average values in bright state presented in Table\,\ref{t3}.

To reconstruct the SED beyond 10\,$\mu$m we used observation of \r\
with IRAS on 12-13.09.1983 (Rao \& Nandy \cite{rao86}).
The $L$ flux of the star on these dates was
smaller than its average value ($\Delta L \simeq 0.25$). To account
for this difference the fluxes measured by IRAS at 12, 25, 60 and
100\,$\mu$m were increased by 8\%, taking into account that $\Delta
f(3.5\,\mu{\rm m})/\Delta f(11\,\mu{\rm m})\approx 3$ (Forrest \e
\cite{forrest72}).

Integrating the SED presented in Fig.\,\ref{f3} we evaluate the
average bolometric flux of \r\ during its bright state as $F_{\rm
bol, mean} \approx 1.55 \times 10^{-7}\,{\rm
erg\,cm^{-2}\,s^{-1}}$. About 85\% of the radiation detected from
the star is emitted within the interval $0.36\,\mu$m\,$\la \lambda
\la 5\,\mu$m, which is completely covered by our photometric
observations. The contributions at $\lambda \la 0.36\,\mu$m,
$\lambda \ga 5\,\mu$m and $\lambda \ga 12\,\mu$m are 4\%, 10\% and
3\%, respectively. In particular, this indicates that the possible
errors in the correction of the IRAS data cannot significantly
change the estimates of the bolometric flux.

   \subsection{Infrared excess}

The strong infrared excess (i.e. the bolometric flux of the dust shell)
can be evaluated as $F_{\rm dust} = F_{\rm bol} - F_{\rm *, bol}$.
Here $F_{\rm bol}$ is the observed bolometric flux and $F_{\rm *,
bol}$ is the bolometric flux emitted by the central star and
passed through the dust shell.

To evaluate the infrared excess we have to reconstruct the
observed SED of the star ($F_{*}(\lambda)$). The upper limit of
the contribution of the dust shell in the $J$ band can be derived
directly from our observations in 1996. In this year a deep
decline was observed ($\Delta J \geq 3.6$), while its $L$
brightness was close to the average value presented in
Table\,\ref{t3}. Attributing the observed $J$ flux in minimum to
the dust shell we conclude that its contribution to the total flux
out of minimum is smaller than 4\%. This indicates that within the
interval $\lambda \leq 1.25\,\mu$m the total flux is approximately
equal to $F_{*}(\lambda)$.

To evaluate $F_{*}(\lambda)$ in $HKLM$  we use the intrinsic
colors of \r\ stars calculated by Asplund \e (\cite{asp97}). They
depend on the effective temperature and the gravitational
acceleration on the star surface. For \r\ we take $T_{\rm eff}
\approx 6750$\,K, $L_{\rm bol} \approx 10^4 L_{\sun}$, $\log{g}
\approx 0.5$, radius $R_* \approx 73.4R_{\sun}$, and
correspondingly the intrinsic colors $(H-K)_0 \approx 0.07$,
$(K-L)_0 \approx 0.06$ and $(L-M)_0 \approx 0.03$ (Asplund \e
\cite{asp97}; Asplund \cite{asp00}). These colors were corrected
for the interstellar and circumstellar reddening although both of
these corrections have only a minor effect on the evaluation of
$F_{\rm*,bol}$. For $\lambda \geq 4.8\,\mu$m we approximate
$F_{*}(\lambda)$ by the blackbody spectrum with the effective
temperature 6750\,K.

The derived SED of the star is presented in Fig.\,\ref{f3}
(long-dashed line). Integrating $F_{*}(\lambda)$ we find $F_{\rm
*, bol} \approx 1.17 \times 10^{-7}\,{\rm erg\,cm^{-2}\,s^{-1}}$
and $F_{\rm dust, mean} \approx 3.8 \times 10^{-8}\,{\rm
erg\,cm^{-2}\,s^{-1}}$. This means that about 24\% of the stellar
radiation is absorbed and re-emitted by the dust shell. Assuming
the geometry of the dust shell to be spherically symmetric, one
can estimate its optical depth as $\tau_{\rm eff, mean} \approx
0.28$. For yellow stars and amorphous carbon grains the value of
the $\tau_{\rm eff,mean}$ differs from the optical thickness at
0.55\,$\mu$m ($\tau$(V)) by a few percent.

   \subsection{Color--brightness diagrams of the star}

   \begin{figure}
      \resizebox{8.5cm}{14cm}{\includegraphics{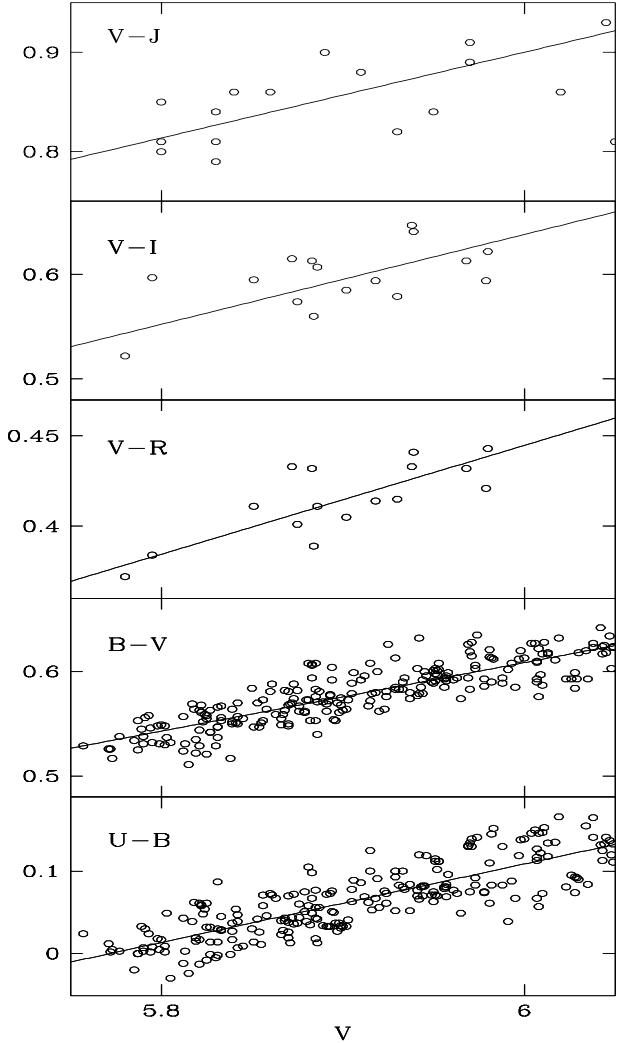}}
         \caption{$(U-B, V)$, $(B-V, V)$, $(V-R, V)$,
$(V-I, V)$ and $(V-J, V)$ diagrams. The solid lines denote the
approximations by linear equations~(\ref{1}--\ref{5})}.
     \label{f4}
    \end{figure}

The derived color--brightness diagrams $(U-B, V)$, $(B-V, V)$,
$(V-R, V)$, $(V-I, V)$ and $(V-J, V)$ are presented in
Fig.\,\ref{f4}. These diagrams have been constructed using the
data from Tables\,1\,and\,2 and the values reported by Fernie
(\cite{fernie82}) and Fernie \e (\cite{fernie86}). For the diagram
$(V-J, V)$ we have chosen the estimates which were obtained in all
three bands on the same date. Only the data relating to the star
brightness in the $L$ band smaller than the average value were
used. The $J$ flux has been reduced by 4\% in order to exclude the
contribution of the dust shell. The linear approximation of the
diagrams yields
  \be\label{1}
U-B = 0.402 V - 2.31,
  \ee
  \be
B-V = 0.280 V - 1.08,
  \ee
  \be
V-R = 0.301 V - 1.36,
  \ee
  \be
V-I = 0.430 V - 1.94,
  \ee
  \be\label{5}
V-J = 0.418 V - 1.62.
  \ee
where $5.79 \leq V \leq 6.05$. It follows from these equations
that the star reddens as its light declines.

    \section{Modeling of the warm dust shell at its mean
brightness in bright state}

Using the quasi-diffusion method of Leung (\cite{leung76}), as
implemented in the CSDUST3 code (Egan \e \cite{egan88}), we have
constructed radiative transfer models which match the SED
of \r. The input parameters of the models were the same
as described by Yudin \e (\cite{yudin01}).

As a first approximation we have assumed that the warm dust shell
is spherical and consists of amorphous carbon grains with the
radius $a_{\rm gr} = 0.01\,\mu$m. The efficiencies of absorption
and scattering of the dust grains were calculated using the Mie
theory. The refractive index was taken from J\"ager \e
(\cite{jag98}, sample cel1000). The intrinsic SED of the star
emission, $F_{*, 0}(\lambda)$, was calculated as $F_{*,
0}(\lambda) = F_*(\lambda)e^{\tau(\lambda)}$, where
$\tau(\lambda)$ is the total optical depth at a given reference
wavelength. The reference wavelength was chosen to be
0.55\,$\mu$m, and the model SED is normalized to the observed one
in the $V$ band.

We have derived the following parameters of the dust shell which
refer to its mean brightness in bright state. The optical
depth at 0.55\,$\mu$m and 2.2\,$\mu$m are $\tau(V) \approx 0.32$ and
$\tau(K) \approx 0.05$, respectively. The inner radius of the dust
shell is $r_{\rm in} \approx 110\,R_*$. The temperature and the
dust density at the inner radius of the dust shell are $T_{\rm
dust}(r_{\rm in}) \approx 860$\,K and $\rho_{\rm dust}(r_{\rm in})
\approx 1.1\times 10^{-20}\,{\rm g\,cm^{-3}}$, respectively, and the
density distribution of the dust shell is $\rho(r) \propto
r^{-2}$.

Using these parameters we have evaluated the dust formation rate
at the inner boundary of the dust shell as $\dot{M}_{\rm dust}
\approx 3.1 \times 10^{-9}\,{\rm M_{\sun}\,yr^{-1}}$. If one
assumes that almost all carbon is bound in the dust grains and
uses the chemical composition of the atmosphere of \r\ from
Asplund \e (\cite{asp00}), the gas density at the inner radius of
the dust shell is $\rho_{\rm gas}(r_{\rm in}) \approx 7.3 \times
10^{-19}\,{\rm g\,cm^{-3}}$. Correspondingly, with a stellar wind
velocity of $V_{\rm env} \approx 45\,{\rm km\,s^{-1}}$, which is
typical for stars of this type, one finds a mass outflow rate in
\r\ of $\dot{M}_{\rm gas} \approx 2.1 \times 10^{-7}\,{\rm
M_{\sun}\,yr^{-1}}$. If, however, a fraction of the carbon and
almost all of the oxygen are bound in the CO molecule, the
corresponding estimates of  $\rho_{\rm gas}(r_{\rm in})$ and
$\dot{M}_{\rm gas}$ turn out to be larger by a factor of 2.5.

As follows from our modeling, the visual brightness of the dust
shell is $V \approx 14\fm6$ and its color indices are $U-B \approx
-0.65$, $B-V \approx -0.28$, $V-R \approx -0.41$ and $R-I \approx
1.47$. The SED of the dust shell is plotted in Fig.\,\ref{f3}
(short-dashed line). The radiation of the dust shell in $UBVR$ is
dominated by scattering. Since the radius of the grains is
significantly smaller than the wavelength, the scattering is
governed by the Rayleigh law ($\sigma_{\rm sca} \propto
\lambda^{-4}$). This explains the blue colors of the dust shell in
this part of the spectrum. In the $I$ band the thermal emission of
the grains begins to dominate over the scattering.

The visual magnitude of \r\ observed during the deepest declines
is about $14^{m}$. Thus, the visual magnitude of the dust shell
should be fainter than $14^{m}$. This was the reason for choosing
a small radius of the dust grains ($\sim 0.01\,\mu$m) in our
calculations. If one assumes that the radius of the dust grains is
larger, the visual brightness of the dust shell becomes greater
than that of the star brightness in deep minima. For instance,
increasing the radius of the dust grains by a factor of 2.5 (i.e.
to 0.025\,$\mu$m), one finds a visual magnitude of the dust shell
of $V \approx 11\fm5$. This reflects the fact that the albedo of
dust grains rapidly increases with their radius.

We have also found the colors of the dust shell to be sensitive to
the value of $r_{\rm in}$. Variations of this parameter of only
10\% lead to variations of the IR excess color indices $(H-K)_{\rm
dust}$ and $(K-L)_{\rm dust}$ by $\sim 0\fm1$. At the same time
these indices are observed to be remarkably stable in contrast to
the brightness of the dust shell (see Sect.\,5). In the context of
our model this means that the value of the inner radius $r_{\rm
in}$ is a fairly constant, time-independent characteristic of the
extended warm dust shell. It should be noted, however, that the
absolute value of $r_{\rm in}$ depends on the optical properties
of the dust grains included in the model calculations. If we used
the extinction of amorphous carbon obtained by Colangeli \e
(\cite{colan95}) (ACAR sample) instead of cel1000 (J\"ager \e
\cite{jag98}), the value of $r_{\rm in}$ would be smaller by a
factor of $\sim 1.4$.

Finally, according to Reimers equation (Reimers \cite{reim75}),
the average mass loss by a yellow supergiant can be estimated as
   \be
\dot{M} = 4 \times 10^{-13}\ \eta\ \frac{L_*  R_*}{M_*}\ {\rm
M_{\sun}\,yr^{-1}},
  \ee
where $L_*$, $R_*$ and $M_*$ are the luminosity, the radius and
the mass of the star expressed in the solar units, and the
dimensionless parameter $\eta$ is limited to $0.3 \leq \eta \leq
3$. Using the corresponding parameters of \r\, we find $\dot{M}
\approx 4.2 \times 10^{-7} \eta {\rm M_{\sun}\,yr^{-1}}$. This is
in agreement with the mass outflow rate derived from our model
provided $\eta \approx 0.5$. Therefore, the mass loss rate of \r\
agrees with the Reimers equation and no superwind is required for
the interpretation of the dust shell formation.

The large value of $r_{\rm in} \approx 110\,R_*$, which does not
change significantly with time, suggests that the extended warm
dust shell of \r\ originates in a similar way as in the case of
ordinary supergiants, i.e. it is caused by the stellar wind and at
a large distance from the star. In addition to this common type of
dust shell of supergiants, the circumstellar shell of \r\ contains
dust clouds which condense somewhere in the vicinity of the star
from high density gas clumps (Feast \cite{fea97}). These
additional dust clouds can explain the deep minima in the visual
light curve. If the dust clouds are considered as an additional
source of the IR emission, then the effective size of this source
is much smaller then the size of the above extended dust shell,
which condenses from the stellar wind. Until now all efforts to
pick out the emission of this consortium of the dust clouds from
the total IR excess were unsuccessful (Feast \cite{fea97}).

   \section{Variability of the warm dust shell}

   \subsection{The dust shell during maximum and
   minimum brightness}

The $UBVRIJHKLM$ magnitudes of \r\ during its maximum (in 1994
April--May) and minimum (in 1999 May--June) brightness in $L$ are
presented in the second and third columns of Table\,\ref{t3},
respectively. The corresponding bolometric flux and infrared
excess of \r, and the optical thickness of the dust shell derived
using these data are presented in Table\,\ref{t4}.

\begin{table}[t]
 \caption[]{The bolometric flux, the infrared excess and the optical
thickness of the dust shell of \r\ during maximum and minimum
of $L$ brightness}
   \label{t4}
\begin{tabular}{cccc}
  \noalign{\smallskip}
  \hline
  \noalign{\smallskip}
 & $F_{\rm bol}$ & $F_{\rm dust}$ & $\tau_{\rm eff}$ \\
 & ${\rm erg\,cm^{-2}\,s^{-1}}$ & ${\rm erg\,cm^{-2}\,s^{-1}}$ &  \\
  \noalign{\smallskip}
  \hline
  \noalign{\smallskip}
Maximum    & $1.90 \times 10^{-7}$ & $6.1 \times 10^{-8}$ & 0.39 \\
Minimum    & $1.24 \times 10^{-7}$ & $2.1 \times 10^{-8 }$ & 0.19 \\
  \noalign{\smallskip}
  \hline
\end{tabular}
\newline
\end{table}

As follows from Table\,\ref{t4}, the amplitude of variations of
the optical thickness of the dust shell in \r\ is $\Delta \tau
\approx 0.2$. At the same time, the value of the color index
$(K-L)_{\rm dust}$ of the IR excess, estimated during maximum and
minimum brightness in the $L$ band, proves to be the same:
$(K-L)_{\rm dust} \approx 2\fm35$. This indicates that the
variations of the optical depth of the dust shell cannot be
explained by the variations of its inner radius, but  reflect the
changes of the dust formation rate at a certain distance from
the star.

  \subsection{Colors and the optical thickness of the dust shell}

    \begin{figure}
      \resizebox{8.5cm}{14cm}{\includegraphics{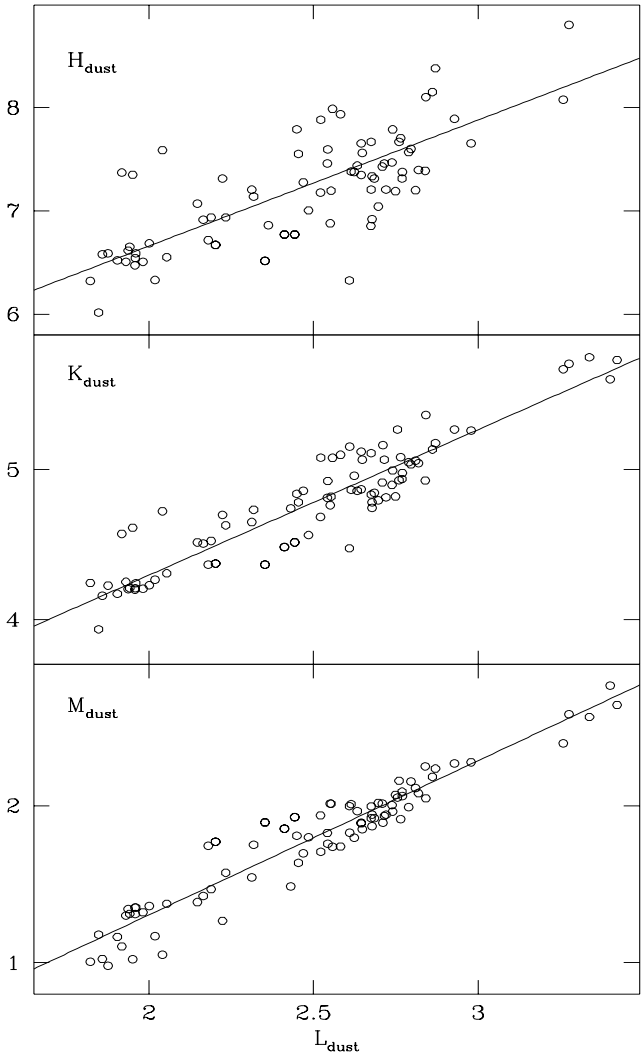}}
          \caption{The $(H_{\rm dust}, L_{\rm dust})$, ($K_{\rm dust},
L_{\rm dust}$) and ($M_{\rm dust}, L_{\rm dust}$) diagrams of the
emission of the dust shell. The solid lines denote the
approximations by linear equations~(\ref{h}--\ref{m})}.
     \label{f5}
   \end{figure}

Fig.\,\ref{f5} shows the color diagrams $(H_{\rm dust}, L_{\rm
dust})$, ($K_{\rm dust}, L_{\rm dust}$), and ($M_{\rm dust},
L_{\rm dust}$). The dots in these diagrams denote IR observations
of \r\ when the star was in its bright state. It was assumed that
the contribution of the dust shell to the flux in $J$ band is
negligibly small. The stellar magnitudes in $HKLM$ and the
infrared excess have been derived using the method described
above. The linear approximation of the diagrams gives
   \be\label{h}
H_{\rm dust} = 1.22 L_{\rm dust} +4.22,
   \ee
   \be
K_{\rm dust} = 0.97 L_{\rm dust} +2.35,
   \ee
   \be\label{m}
M_{\rm dust} = 0.98 L_{\rm dust} -0.66.
   \ee
For the mean $L$ brightness, $L_{\rm dust, mean} \approx 2.49$, we find
$(H-L)_{\rm dust, mean} = 4.76 \pm 0.4$, $(K-L)_{\rm dust, mean} =
2.28 \pm 0.15$, and $(L-M)_{\rm dust, mean} = 0.70 \pm 0.12$.

As follows from Eqs.~(\ref{h}-\ref{m}), the variations of the dust
shell brightness are not accompanied by  remarkable variations of its
colors. Hence, there is a proportionality between the
bolometric stellar magnitude of the dust shell, $m_{\rm dust, bol}$,
and its $L$ magnitude:
  \be\label{mb}
m_{\rm dust, bol} \simeq L_{\rm dust} + 4\fm 55.
  \ee

We can calculate the function $\tau_{\rm eff} = f(F_{\rm dust})$
from the expression $F_{\rm dust} = F_{\rm bol}(1-exp(-\tau_{\rm
eff}))$. Finally, using a quadratic polynomial approximation of
the functions $L_{\rm dust}=f(L)$ for the data of our IR
observations, we find:
  \be
m_{\rm dust, bol} \simeq 0.046 L^2 + 0.88 L +4.68,
  \ee
  \be\label{last}
\tau_{\rm eff} \simeq 0.176 L^2 - 1.20 L +2.14.
  \ee
The function $\tau_{\rm eff}(t)$ for the period 1994--2001 is
plotted in Fig.\,\ref{f6}. Our modeling has shown that such a
sharp transition from the decrease to the increase of the optical
depth requires an abrupt change of the dust formation rate close
to the minimum. This means that there was a sharp density
enhancement in the stellar wind. The solid line in Fig.\,\ref{f6}
shows the function $\tau_{\rm eff}(t)$, calculated with the
assumption that at minimum the rate of dust formation sharply
increases in the minimum of $\tau_{\rm eff}$ by a factor of 4.3,
and then rises $\propto t^{1.3}$ during about 560\,days until the
maximum value of the optical depth is reached.

The dust formation rate depends on the gas density at the inner
boundary of the dust shell and, therefore, on the intensity of the
mass-loss. The time lag between the increase of the stellar wind
and the increase of gas density at the inner radius of the shell
is
    \be
t_{\rm lag} = \frac{r_{\rm in}}{V_{\rm env}} \simeq 4\
\left(\frac{r_{\rm in}}{110\,{\rm R_*}}\right)\ \left(\frac{V_{\rm
sw}}{45\,{\rm km\,s^{-1}}}\right)\ {\rm yr}.
   \ee

Interestingly, this 4-year wind travel time can probably be seen
in the $V$ and $L$ light curves. The comparison of the $V$ and $L$
light curves shows that the period 1994--1999 of decreasing $L$
and $M$ brightness follows approximately 4 years after the period
of constant V magnitude between 1989 and 1995. We suggest the
following explanation for this. We assume that during phases of
increased stellar activity there is a simultaneous increase of
both (1)\,the stellar (supergiant) wind and (2)\,the rate of the
formation of (eclipsing) dust clouds (and we assume that the
eclipsing dust clouds cause the deep minima in the $V$ light
curve). If these two effects are correlated, then 1983--1989 (with
several $V$ minima) was a period of increased activity and,
therefore, increased stellar wind. After the wind travel time of
approximately 4 years, this increased stellar wind arrived at the
extended dust shell region and then caused the $L$ band brightness
increase. On the other hand, in the period 1993--1999 the $L$ band
brightness was decreasing because in this period only lower
intensity stellar wind arrived, which was produced during the
low-activity period 1989--1995 (high $V$ brightness).

   \begin{figure}
      \resizebox{8.5cm}{8cm}{\includegraphics{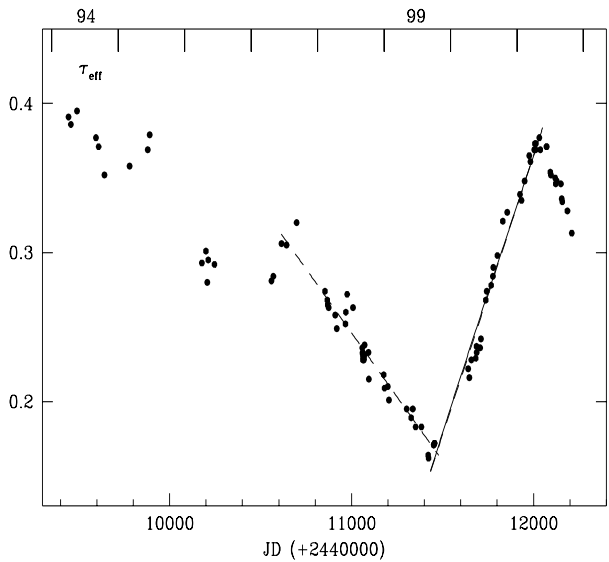}}
         \caption{Optical thickness {$\tau_{\rm eff}$ of the dust
shell versus time (see text)}}.
     \label{f6}
   \end{figure}

    \subsection{Periodicity}

In order to search for the periodicity in the variations of the
dust shell brightness, i.e. of the mass-loss rate, the $L$ light
curve has been analyzed using the code developed by
Yu.K.\,Kolpakov (\mbox{http://infra.sai.msu.ru/prog/kolpakov}). We
combined our observations with the data presented by Strecker
(\cite{streck75}; 35 points during 1968--1974) and Feast et al.
(\cite{feast97}; 24 points during 1983--1991), and removed the
points obtained during visual declines. Two harmonics of
semi-regular variations of the star brightness in the $L$ band
were derived: $P_1 \simeq 1206$\,days, $\Delta L_1 \approx 0\fm8$
and $P_2 \simeq 4342$\,days, $\Delta L_2 \approx 0\fm6$. The
corresponding phase diagrams are presented in Fig.\,\ref{f2}. The
dates of minima of these harmonics are given by the following
expressions:
   \be\label{m14}
{\rm min}(L_1) = 2442986 + 1206^d {\rm E},
   \ee
   \be\label{m15}
{\rm min}(L_2) = 2442860 + 4342^d {\rm E}.
   \ee
These dates are marked by vertical bars on Fig.\,\ref{f1}. The
minima of these harmonics coincided in 1999, when the star
brightness in $L$ and $M$ was in unusually deep minimum. It should
be noted that the existence of the two dominant modes superimposed
with irregular variations makes the $L$ light curve quite complex.
Moreover, the shape of the harmonics differs from the simple sine
wave shown in Fig.\,\ref{f2}. For example, the rise of the $L$
brightness in 1984 and 1988 was rather sharp compared to the
subsequent declines (Fig.\,\ref{f1}). The values of the derived
periods are very close to those of the mean time interval between
the individual visual declines of \r, $\sim 1100$\,days (Jurcsik
\cite{jur96}), and between groups of visual minima, $\sim
4400$\,days (Rosenbush \cite{ros01}).

In the bright state of \r, the $L$ variability is mainly caused by
changes of the optical depth of the shell. Hence, if the dust
shell is spherical, we would expect to find the corresponding
semi-regular variations in the visual star brightness with period
$P_1 \approx 1206$\,days, which are anti-phased to those in the
$L$ band. According to Eq.~(\ref{m14}) their full amplitude would
be $\Delta V \approx 0.3$\,mag. However, a period search has not
revealed this harmonic in the $V$ light curve. This indicates that
the density of the stellar wind (and, correspondingly, the shell
density) along the line of sight is smaller than the average one,
i.e. the dust shell is asymmetric.

In this case the derived optical depth of the dust shell can be
considered as the sphere-averaged one. If the dust shell is
optically thin, then its bolometric flux is proportional to the
optical depth averaged over the sphere. Since the shell's SED is
primarily defined by the value of the inner radius (i.e., $T_{\rm
dust}(r_{\rm in})$) our above results concerning the inner radius
(its size and time-independence) and mass-loss rate would be
valid.

    \section{Concluding remarks}

We have presented optical and IR long-term monitoring and
radiative transfer modeling of \r. These studies have allowed the
derivation of various time-dependent properties of both (i)\,the
star itself at bright state and (ii)\,an extended dust shell
(similar to that observed around many supergiants) which condenses
from the supergiant wind at a large distance of approximately 100
stellar radii from the star. This extended dust shell is larger
than the suggested region of dust clouds (distance approx. 2--30
stellar radii, see, e.g. Clayton \cite{clay96}, Fadeyev
\cite{fad88}), which causes the deep minima in the visual light
curve. The extended dust shell (radius $\sim 19$\,mas) was first
resolved by speckle interferometric observations with the SAO 6\,m
telescope (Ohnaka \e \cite{ohnaka01}).

In the $V$ and $L$ light curves probably a $\sim$\,4-year wind
travel time from the stellar surface to the extended dust shell
with a radius of approximately 110 stellar radii can be seen. The
comparison of the $V$ and $L$ light curves shows that the period
1994--1998 of an increased (and decreasing) $L$ and $M$ brightness
started  approximately 4 years after the end (in 1990) of a period
with many deep minima in the visual light curve. This can be
explained if we assume that during phases of increased stellar
magnetic activity (see studies by Soker \& Clayton \cite{soker99})
there is a simultaneous increase of both (i)\,the supergiant wind
and (ii)\,the rate of formation of dust clouds. If this assumption
is true, the increased stellar wind period ended in 1990. The L
band brightness increased until 1994. This can be explained by the
increased stellar wind until 1990 and a wind travel time of 4
years from the stellar surface to the extended dust shell region.
In the period 1994-1999 the L band brightness was decreasing
because in this period only lower intensity stellar wind arrived,
which was produced during the low-activity period 1990-1995. We
believe that the variation of the stellar wind can be considered
as an indirect argument in favor of active phenomena on the
surface of \r.

\begin{acknowledgements}
Nazar Ikhsanov  acknowledges the support of the Alexander von
Humboldt Foundation within the Long-term Cooperation Program.
This research was partly supported by Russian Foundation for
Basic Research (project No 02-02-16235).
\end{acknowledgements}

\end{document}